\documentclass[3p,a4paper,twocolumn]{elsarticle}

\usepackage{amsfonts}
\usepackage{amsmath}
\usepackage{amssymb}
\usepackage{graphicx}%
\begin{document}
\title{Quantum computation in continuous time using dynamic invariants}
\author{M. S. Sarandy\corref{cor1}\fnref{fn1}}
\ead{msarandy@if.uff.br}
\address{Instituto de F\'{\i}sica, Universidade Federal Fluminense,
Av. Gal. Milton Tavares de Souza s/n, Gragoat\'a, 24210-346, Niter\'oi, RJ, Brazil.}
\author{E. I. Duzzioni}
\ead{duzzioni@infis.ufu.br}
\address{Instituto de F\'{\i}sica, Universidade Federal de Uberl\^{a}ndia, Caixa
Postal 593, 38400-902, Uberl\^{a}ndia, MG, Brazil}
\author{R. M. Serra}
\ead{serra@ufabc.edu.br}
\address{Centro de Ci\^{e}ncias Naturais e Humanas, Universidade Federal do ABC, R.
Santa Ad\'{e}lia 166, Santo Andr\'{e}, 09210-170, S\~{a}o Paulo, Brazil}
\cortext[cor1]{Corresponding author}
\fntext[fn1]{Tel.: +55-21-2629-5802 / Fax: +55-21-2629-5887 }
\begin{abstract}
We introduce an approach for quantum computing in continuous time based on 
the Lewis-Riesenfeld dynamic invariants. This approach allows, under certain 
conditions, for the design of quantum algorithms running on a nonadiabatic regime.  
We show that the relaxation of adiabaticity can be achieved by processing information in the
eigenlevels of a time dependent observable, namely, the dynamic invariant
operator. Moreover, we derive the conditions for which the computation can be
implemented by time independent as well as by adiabatically varying
Hamiltonians. We illustrate our results by providing the implementation of
both Deutsch-Jozsa and Grover algorithms via dynamic invariants.
\end{abstract}

\maketitle


\section{Introduction}

Quantum information processing can be implemented through different quantum
computation (QC) models. One promising such a model is provided by adiabatic
QC (AQC)~\cite{Farhi:01}. In AQC, rather than using a circuit of unitary
quantum gates as in the standard QC model (SQC), an algorithm is implemented
via the slow continuous evolution of a time-dependent Hamiltonian $H(t)$. The
quantum system is prepared in some simple eigenstate $|n(0)\rangle$ of the
initial Hamiltonian $H(0)$ and is then allowed to evolve adiabatically so that
it remains in the corresponding instantaneous eigenstate $|n(t)\rangle$ of
$H(t)$ at all times. At the end of the process, the solution of the problem is
encoded in the final state of the system, whence it can be read out by means
of a convenient measurement. 
Protection of AQC against decoherence has been investigated in several
works~\cite{AQC:decoh,Sarandy:05,Lidar:08},
settling AQC as a favorable approach for QC in real (open) quantum systems.

However, while decoherence-protected AQC is potentially attainable~\cite{Lidar:08}, 
adiabatic steps may be a harsh requirement in several experiments~\cite{Chen:10}. 
Moreover, nonadiabatic shortcuts are also helpful to clarify the role played 
by adiabaticity for QC in continuous time. In this context, inspired by AQC, 
the aim of this work is to propose an alternative approach to perform
QC via continuous evolution in Hilbert space, which is based on the theory of
dynamic invariants introduced by Lewis and Riesenfeld~\cite{Lewis:67,Lewis:69}. 
The theory of dynamic invariants was conceived as a tool to solve time-dependent 
problems in quantum mechanics. In turn, as a first application, it
was used to discuss the nonadiabatic dynamics of a time-dependent harmonic
oscillator~\cite{Lewis:67}. Since then, the dynamic invariants technique has
been applied to a number of problems, which include quantum
optics~\cite{Villas-Boas 2003}, atomic
systems~\cite{Mizrahi:90}, and geometric phases~\cite{Sarandy:07,Duzzioni:08}. 

In the present work, we will show that dynamic invariants can be used to implement a
nonadiabatic approach to perform QC. In QC by dynamic invariants (QCDI), the
computation process will be developed in an arbitrary eigenstate (here chosen
as the lowest eigenvalue state) of a time-dependent quantum observable -- the so-called
dynamic invariant operator, which will conveniently be defined below. 
The final (target) state is achieved in a nonprobabilistic way and 
the procedure is independent of the adiabatic approximation. 
Nevertheless, QCDI is not proposed here to supersede 
the adiabatic approach, since the required unitary interpolation 
for the dynamic invariant operator may lead to the necessity of 
many-body interactions in the Hamiltonian, whose simulation compromises
scalability. However, the method provides a suitable implementation 
of a quantum algorithm in continuous time either if nonadiabaticity is 
needed in a test-bed small-scale QC or if many-body interactions 
can be avoided in a particular problem. 
As an illustration of QCDI, we will provide implementations 
for both Deutsch-Jozsa and Grover algorithms. 


\section{Lewis-Riesenfeld dynamic invariants}


\label{SecDI}

For a closed quantum system, a dynamic invariant $I(t)$ is defined as an Hermitian
operator that satisfies~\cite{Lewis:67,Lewis:69}
\begin{equation}
\frac{\partial I(t)}{\partial t}+\frac{i}{\hbar}\left[  H(t),I(t)\right]
=0,\label{diclosed}%
\end{equation}
where $H(t)$ is the Hamiltonian of the system and, from now on, $\hbar$ will be
set to one. Dynamic invariants are quantum mechanical constants of motion,
implying therefore that their expectation values are constant, i.e., $d\langle
I(t)\rangle/dt=0$. The construction of such an operator allows for the direct
integration of the time-dependent Schr\"{o}dinger equation
\begin{equation}
H(t)|\psi(t)\rangle=i|{\dot{\psi}}(t)\rangle,\label{se}%
\end{equation}
with the dot symbol denoting time derivative. Let us consider an instantaneous
orthonormal eigenbasis for $I(t)$
\begin{equation}
I(t)|\varphi_{i}(t)\rangle=\lambda_{i}|\varphi_{i}(t)\rangle,\label{Ibasis}%
\end{equation}
where we assume, for simplicity, that the $I(t)$ has non-degenerate
eigenlevels. Then, we expand the wave function $|\psi(t)\rangle$ in the
invariant operator basis $\{|\varphi_{i}(t)\rangle\}$, yielding
\begin{equation}
|\psi(t)\rangle=\sum_{i}c_{i}(t)|\varphi_{i}(t)\rangle.\label{psiexp}%
\end{equation}
By inserting Eq.~(\ref{psiexp}) in Eq.~(\ref{se}) and projecting the result
onto $\langle\varphi_{j}(t)|$ we obtain
\begin{equation}
{\dot{c}}_{j}=-\sum_{i}c_{i}\left(  i\langle\varphi_{j}|H|\varphi_{i}%
\rangle+\langle\varphi_{j}|{\dot{\varphi}}_{i}\rangle\right)  .\label{step1}%
\end{equation}
On the other hand, taking the derivative of Eq.~(\ref{Ibasis}) and projecting
it onto $\langle\varphi_{j}(t)|$ we get
\[
\dot{\lambda}_{i}\delta_{ij}+\left(  \lambda_{i}-\lambda_{j}\right)  \left(
i\langle\varphi_{j}|H|\varphi_{i}\rangle+\langle\varphi_{j}|{\dot{\varphi}%
}_{i}\rangle\right)  =0.
\]
The equation above implies that
\begin{align}
i &  =j:\dot{\lambda}_{i}=0\Rightarrow\lambda_{i}={\text{constant}%
},\label{cond1}\\
i &  \neq j:\langle\varphi_{j}|{\dot{\varphi}}_{i}\rangle=-i\langle\varphi
_{j}|H|\varphi_{i}\rangle.\label{cond2}%
\end{align}
Equation~(\ref{cond1}) is a direct consequence of $I(t)$ being a constant of
motion. Concerning Eq.~(\ref{cond2}), it allows for the integration of
Schr\"{o}dinger equation. Indeed, use of Eq.~(\ref{cond2}) into
Eq.~(\ref{step1}) yields
\[
c_{j}(t)=c_{j}(0)e^{\left[  -\int_{0}^{t}d\tau\left(  \langle\varphi
_{j}|\frac{\partial}{\partial\tau}|\varphi_{j}\rangle+i\langle\varphi
_{j}|H|\varphi_{j}\rangle\right)  \right]  }.
\]
Therefore, if we initially prepare the system in the eigenstate $|\varphi
_{j}(0)\rangle$ of $I(t)$ then the system will necessarily evolve to
$|\varphi_{j}(t)\rangle$ at any time $t$. The nontransitional evolution of an
eigenstate of $I(t)$ plays the role of the adiabatic evolution of an
eigenstate of $H(t)$. As we will show, this
suitably built evolution in Hilbert space can be used to perform QC with no
adiabatic constraint.


\section{Quantum computation by invariants}


\label{SecQCDI}

Let us now introduce a mechanism to perform QCDI. Our approach, proposed here 
to implement QC, closely resembles in several aspects the invariant-based inverse 
engineering method to accelerate adiabatic processes via nonadiabatic 
shortcuts~\cite{Muga:09,Muga:10,Chen-Review,Muga:11} as well as the Berry's 
transitionless tracking algorithm~\cite{Berry:09}. 

First, before the definition of the Hamiltonian operator $H$, we introduce a 
time-dependent dynamic invariant $I(s)$, where $s$ denotes the normalized time, namely, 
$s=t/T$, with $T$ standing for the total evolution time and $0\leq s\leq1$. The operator
$I(s)$ is constructed such that: (a) $I(0)$ has a nondegenerate lowest eigenvalue state $|\phi(0)\rangle$ 
exhibiting a simple structure; (b) $I(1)$ has a nondegenerate lowest eigenvalue state $|\phi(1)\rangle$
that contains the solution of the problem (similarly to AQC, this can be
obtained by providing an eigenvalue penalty for any state that violates the
solution to be found); (c) $I(s)$ is defined, for intermediary values of $s$
$(0<s<1)$, by a conveniently chosen interpolation. Just for simplicity, we will call  
the lowest eigenvalue state of $I(s)$ from now on as its ground state.

\begin{figure}
\includegraphics[height=4.6 cm,width=6.5 cm]{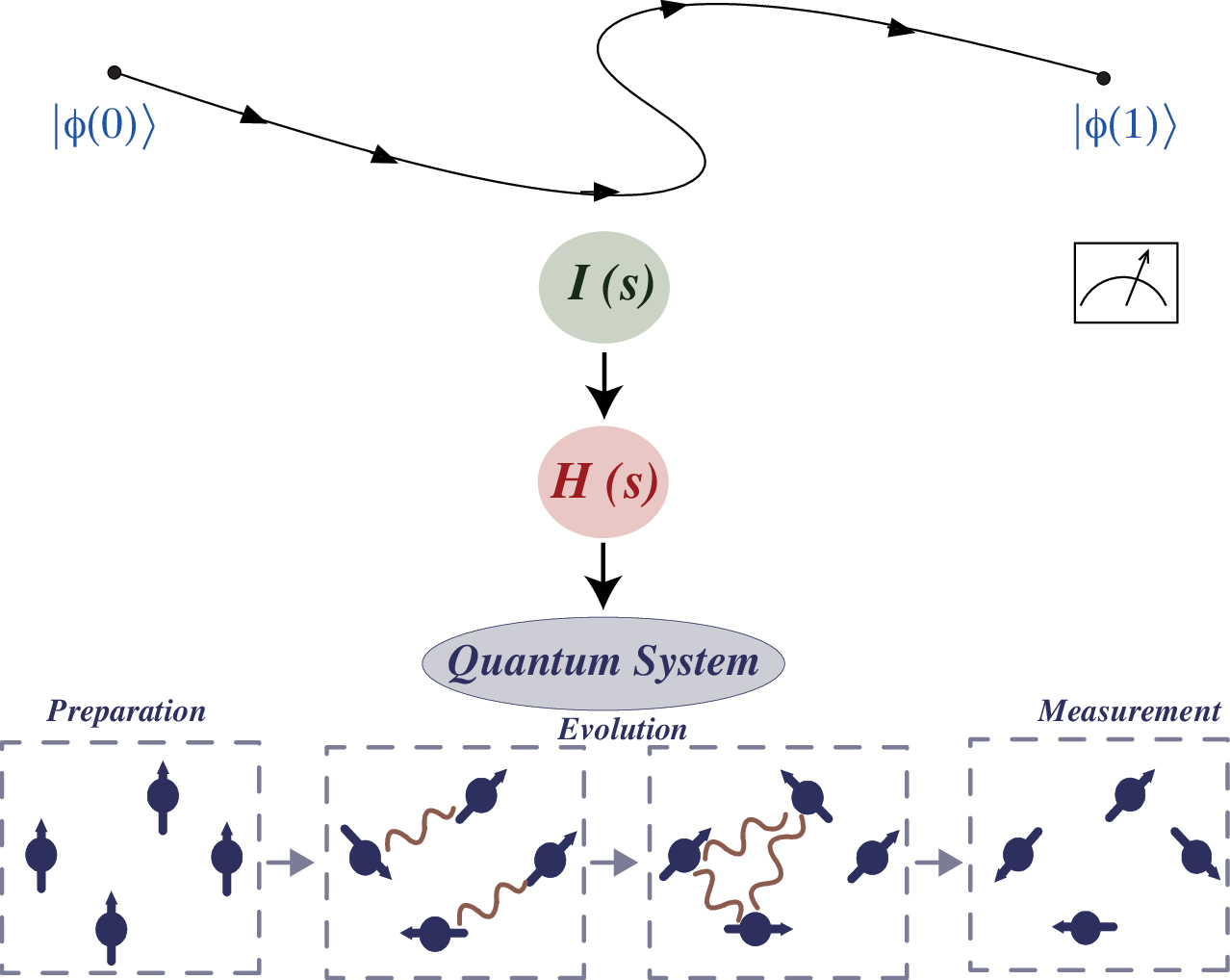}
\caption{(Color online) Schematic description of QCDI. An evolution in continuous time connects the initial and final 
lowest eigenvalue states of the interpolating dynamic invariant operator. The invariant operator then fixes the 
Hamiltonian to be implemented through a suitable quantum system. }
\label{f1}
\end{figure}

As a second step, we can
determine the Hamiltonian under which the system will be evolved by requiring
that $I(s)$ is a dynamic invariant. This is done here after the definition of
$I(s)$ and can be achieved by imposing Eq.~(\ref{diclosed}) which, in terms of
the normalized time $s$, becomes
\begin{equation}
\frac{\partial I(s)}{\partial s}+iT\left[  H(s),I(s)\right]  =0.
\label{dicloseds}%
\end{equation}
As a final step, we prepare the system in the ground state $|\phi(0)\rangle$ of
$I(0)$ and let it evolve during a fixed evolution time $T$. The system will
then be naturally led to the corresponding ground
state $|\phi(1)\rangle$ of $I(1)$, since $I(s)$ is built (by
definition) as a dynamic invariant. As the solution of the problem is encoded
in $|\phi(1)\rangle$, then it can be read out from a suitable measurement. 
The correct final state $|\phi(1)\rangle$ is reached with absolute certainty 
in a nonprobabilistic way. A schematic description of QCDI is provided in Fig.~\ref{f1}.  
The form of Eq. (\ref{dicloseds}) leads to a unitary evolution for $I(s)$, i.e.,%
\begin{equation}
I(s)=\widetilde{U}(s)I(0)\widetilde{U}^{\dag}(s), \label{Is}%
\end{equation}
where $\widetilde{U}(s)$ is the unitary evolution operator and $I(0)$ is the 
invariant operator at $s=0$. 
The unitary interpolation for $I(s)$ given in Eq.~(\ref{Is}) has the property of 
preserving the spectral gaps among the eigenvalues of $I(s)$ during
all the evolution~\cite{Siu:04}. This ensures the absence of level crossings
in the spectrum of $I(s)$. 
Note that no adiabaticity constraint is imposed on the evolution of the
invariant operator $I(s)$ and, consequently, on the evolution of the interpolating state 
$|\phi(s)\rangle$. If we allow for an adiabatic evolution of $I(s)$, i.e., by
using that $\partial I(s)/\partial s\simeq0$, we obtain from Eq.~(\ref{dicloseds}) that 
the eigenstates of $I(s)$ become also eigenstates of
the Hamiltonian, since $\left[  H(s),I(s)\right]  \simeq0$. For such a case, 
QCDI gets completely equivalent to AQC.
On the other hand, if we discretize the unitary transformation $\widetilde{U}(s)$ 
as sequence of quantum gates, we recover the SQC model. In this case, 
$\widetilde{U}(s) = \widetilde{U}_1(s) \otimes \cdots \otimes \widetilde{U}_N(s)$, 
where $\widetilde{U}_i(s)$ stands for a one or two-qubit quantum gate.

\section{Interpolation of the dynamic invariant}


\label{SecInter}

We consider now a possible strategy to implement the unitary interpolation of a 
dynamic invariant which evolves from $|\phi(0)\rangle$ to $|\phi(1)\rangle$. 
Let us begin by expanding the unitary evolution for $I(s)$ as
\begin{equation}
{\widetilde{U}}(s)=\exp\left[  i\sum_{i=1}^{N}f_{i}(s)O_{i}\right]  ,
\label{Interpol}%
\end{equation}
with $f_{i}(s)$ being real functions of time and $O_{i}$ time independent
Hermitian operators. To obtain the Hamiltonian that 
implements the evolution operator given in Eq.~(\ref{Interpol}), we can apply 
the theory of dynamic invariants as follows.
Let ${\mathcal{C}_{N}}=\left\{  O_{1},\ldots,O_{N}\right\}  $ be the set of operators
composing $\widetilde{U}(s)$ in Eq. (\ref{Interpol}). We assume that ${\mathcal{C}_{N}}$
is a subset of ${\mathcal{C}_{M}}=\left\{  O_{1},\ldots, O_{M}\right\}  $, with $M\geq N$,
where the elements of ${\mathcal{C}_{M}}$ define an arbitrary Lie algebra 
$\left[  O_{i},O_{j}\right]  =\sum_{k=1}^{M}C_{ij}^{k}O_{k}$. We
write $I(0)$ as
\begin{equation}
I(0)=\sum_{i=1}^{M}\lambda_{i}(0)O_{i}, \label{I00}%
\end{equation}
with $\lambda_{i}(0)$ being real coefficients. Then, substituting
Eqs.~(\ref{Interpol}) and (\ref{I00}) into Eq.~(\ref{Is}), we obtain 
$I(s)=\sum_{k=1}^{M}\lambda_{k}(s)O_{k}$, with
\begin{equation}
\lambda_{k}(s) = \left[  \lambda_{k}(0)+i\sum_{i=1}^{N}\sum_{j=1}^{M} 
f_{i}(s)C_{ij}^{k} \lambda_{j}(0) + \cdots \right].
\end{equation}
As the
operators $O_{k}$ are elements of a Lie algebra, we take the Hamiltonian of
the system as a linear combination of such operators, namely,  
$H(s)=\sum_{k=1}^{M}h_{k}(s)O_{k}$,
with $h_{k}(s)\in%
\mathbb{R}
$. This expansion of $H$ is rather convenient since it ensures that, after
evaluating the invariant operator $I(s)$, we may obtain the coefficients
$h_{k}(s)$ through Eq.~(\ref{dicloseds}). Moreover, note that $h_{k}(s)\,$
can be determined by the solution of a set of coupled linear algebraic equations instead
of a set of linear differential equations. In particular, taking qubits as the
building blocks of QC, we can always expand the Hamiltonian in terms of tensor
product of Pauli spin matrices (satisfying the $su(2)$ algebra) in the form
\begin{equation}
H(s)=\sum_{\left\{  k_{i}\right\}  }h_{1,\ldots,n}^{k_{1},\ldots,k_{n}%
}(s)\sigma_{1}^{k_{1}}\otimes\sigma_{2}^{k_{2}}\otimes\ldots\otimes
\sigma_{n}^{k_{n}},
\label{hgenla}
\end{equation}
where the lower indices enumerate $n$ qubits and the upper indices refer to the
set $\left\{  \mathbf{1},\sigma^{x},\sigma^{y},\sigma^{z}\right\}  $ of
identity and Pauli spin-$1/2$ matrices. The coefficients $h_{i}^{k_{i}%
}(s)\in%
\mathbb{R}
$ since $H(s)$ is Hermitian. The Hamiltonian given by Eq.~(\ref{hgenla}) 
will exhibit many-body interactions if all the coefficients 
$h_{1,\ldots,n}^{k_{1},\ldots,k_{n}}(s)$ are nonvanishing. 
Naturally, the simulation of such a Hamiltonian is typically hard. 
However, as mentioned before, this approach provides a 
suitable implementation either if nonadiabaticity is needed in a test-bed small-scale 
QC or if many-body interactions can be avoided in a particular problem. 


\section{Applications}


\label{SecIllus}


\subsection{Example 1: the Deutsch-Jozsa problem}


Given a binary function $f:\{0,1\}^{n}\rightarrow\{0,1\}$ ($n$ is the number
of bits) which is promised to be either constant or balanced, the Deutsch-Jozsa (DJ) problem
consists in determining which type the function is. Here we construct an
implementation by dynamic invariants for the optimized version of the
algorithm~\cite{Collins:98} 
(see also Refs.~\cite{Sarandy:05,aqc-dj} for AQC formulations for the DJ problem).

\subsubsection{DJ problem for n=1}

Let us begin with the simple case $n=1$. The
input state is $|\phi(0)\rangle=|+\rangle$, where $|\pm\rangle
=(|0\rangle\pm|1\rangle)/\sqrt{2}$, with $\{|0\rangle,|1\rangle\}$ being the
computational basis for the qubit (eigenstates of the Pauli matrix $\sigma
^{z}$). The initial dynamic invariant is chosen such that its ground state is
$|\phi(0)\rangle$, i.e., $I(0)=\omega|-\rangle\langle-|$, where
$\omega$ is a free parameter introduced to set the gap between the eigenstates
of $I(0)$. Note that $I(0)$ is introduced in a such a way that a penalty is
provided for any state having a contribution of $|-\rangle$. Hence
$|\phi(0)\rangle$ is its ground state. The DJ problem can be solved by
a single computation of the function $f$ through the unitary transformation
$U|x\rangle=(-1)^{f(x)}|x\rangle$ ($x\in\{0,1\}^{n}$)~\cite{Collins:98}, so
that in the $\{|x\rangle\}$ (computational) basis $U$ is represented by the
diagonal matrix $U=\mathrm{diag}[(-1)^{f(0)},(-1)^{f(1)}]$. In terms of the
Pauli matrices the operator $U$ may be written as $U=\xi_{+}\mathbf{1}+\xi
_{-}\sigma^{z}$, where $\xi_{\pm}=(1/2)[(-1)^{f(0)}\pm(-1)^{f(1)}]$. 

Our implementation requires a final dynamic invariant $I(1)$ such that its 
ground state is
$|\phi(1)\rangle=U|\phi(0)\rangle=\xi_{+}|+\rangle+\xi_{-}|-\rangle$.
This is accomplished by a unitary transformation on $I(0)$, i.e.,
$I(1)=UI(0)U^{\dagger}$. Note that this is similar to the nonlinear
interpolation for the DJ problem proposed in Ref.~\cite{Sarandy:05} in the
context of AQC. 
However, the nonlinear interpolation is
implemented here on $I(s)$ instead of being realized on $H(s)$. The final
dynamic invariant encodes the solution of the DJ problem in its ground state,
which can be extracted via a measurement of the qubits in the basis
$\{|+\rangle,|-\rangle\}$. Indeed, for a constant function, we obtain $\xi
_{+}=\pm1$ and $\xi_{-}=0$. Then $|\phi(1)\rangle=|+\rangle$ (up to a
possible global phase). On the other hand, for a balanced function, we have
$\xi_{+}=0$ and $\xi_{-}=\pm1$. Then $|\phi(1)\rangle=|-\rangle$ (up to
a possible global phase). In order to explicitly evaluate $I(s)$ we consider 
the evolution operator in the form
\begin{equation}
\widetilde{U}(s)=\exp\left[  -i\alpha(s)U\right] .
\label{Interpol1}
\end{equation}
Then, from Eq.~(\ref{Is}), we obtain
\begin{equation}
I(s)=\frac{\omega}{2}\left\{  \mathbf{1}-\cos\left[  2\alpha(s)\xi_{-}\right]
\sigma^{x}-\sin\left[  2\alpha(s)\xi_{-}\right]  \sigma^{y}\right\}  .
\end{equation}
Since the operator $U$ displays the properties $UU^{\dag}=\mathbf{1}$ and
$U=U^{\dag}$, we can implement the algorithm through the Hamiltonian
\begin{equation}
H(s)=\frac{1}{T}\frac{d\alpha(s)}{ds}U=\frac{1}{T}\frac{d\alpha(s)}%
{ds}\left(  \xi_{+}\mathbf{1}+\xi_{-}\sigma^{z}\right)  .
\end{equation}
Hence, by controlling the time variation of $\alpha(s)$ and the frequency
$T^{-1}$ we can optimize the run time of the algorithm. 
Naturally, the run time is constrained by the quantum 
brachistochrone~\cite{Brody:06}, which poses a physical limitation 
on the speed of unitary transformations.

\subsubsection{DJ problem for n=2}

Let us consider now a possible generalization for the case $n=2$. In this case, 
we can apply a simpler interpolation scheme in comparison with the method delineated 
in Section~\ref{SecInter}. As we will see, since the initial and the final invariant 
operators exhibit similar forms, we just need to replace their coefficients by time-dependent 
funtions obeying the required boundary conditions at $s=0$ and $s=1$. Such strategy 
will allow us to find out the simplest Hamiltonian to implement the algorithm.
We begin by 
taking the initial state of the system as 
$\left\vert \phi\left( 0\right) \right\rangle =\left\vert
+\right\rangle _{1}\left\vert +\right\rangle _{2}$.
A simple initial dynamic invariant $I\left( 0\right)$ that exhibits 
$\left\vert \phi\left( 0\right) \right\rangle$ as its ground state 
can be defined by imposing an eigenvalue penalty for every individual spin 
whose quantum state has a contribution of the basis state $|-\rangle$, i.e.
$I\left( 0\right) =\left\vert -\right\rangle _{1}\left\langle -\right\vert
\otimes 1_{2}+1_{1}\otimes \left\vert -\right\rangle _{2}\left\langle
-\right\vert$.
The final dynamical invariant $I\left( 1\right) $ can be obtained by the
application of the unitary operator
$U= \sum_{i=0}^{3} \left[\left( -1\right) ^{f\left(i\right)}
\left\vert e_i\right\rangle \left\langle e_i\right\vert \right]$,
where $|e_0\rangle = |00\rangle$, $|e_1\rangle = |01\rangle$, 
$|e_2\rangle = |10\rangle$, and $|e_3\rangle = |11\rangle$.
Indeed, we can rewrite $I\left( 0\right)$ in the computational basis as
\begin{eqnarray}
I\left( 0\right) =\mathbf{1}_{12}-\frac{1}{2}\left[ \left( \left\vert
e_0\right\rangle +\left\vert e_3\right\rangle \right) \left( \left\langle
e_1\right\vert +\left\langle e_2\right\vert \right) \right. \nonumber \\
\left. +\left( \left\vert e_1\right\rangle +\left\vert e_2\right\rangle \right) 
\left( \left\langle e_0\right\vert +\left\langle e_3\right\vert \right) \right] .
\end{eqnarray}
Then, by using that $I\left( 1\right) =UI\left( 0\right) U^{\dag }$, we obtain
\begin{eqnarray}
I\left( 1\right) &=& \mathbf{1}_{12}-\frac{1}{2} 
\left[ \left(-1\right)^{f\left(0\right)+f\left(1\right)} \rho_{01} \right. \nonumber \\
&&\hspace{1cm}\left.+\,\left(-1\right) ^{f\left(0\right) +f\left(2\right)}\rho_{02}\right.\nonumber \\ 
&&\hspace{1cm}\left. +\,\left(-1\right) ^{f\left( 1\right) +f\left( 3\right)}
\rho_{13} \right.\nonumber \\ 
&&\hspace{1cm}\left.+\,\left(-1\right)^{f\left( 2\right) +f\left(3\right)}\rho_{23} \right] ,
\end{eqnarray}%
where $\rho_{ij} = \left\vert e_i\right\rangle \left\langle e_j\right\vert
+\left\vert e_j\right\rangle \left\langle e_i\right\vert$.
The ground state of $I(1)$ is
$\left\vert \phi\left( 1\right) \right\rangle = \sum_{i=0}^{3} 
\frac{1}{2}\left[\left(-1\right)^{f\left(i\right) }\left\vert e_i\right\rangle  \right]$.
Note that both the initial and final invariants display the same structure, 
which allows for the definition of the interpolating invariant as
\begin{eqnarray}
\hspace{-0.7cm}I\left(s\right) &=&\mathbf{1}_{12}-\left[ \alpha \left( s\right) \left\vert
e_0\right\rangle \left\langle e_1\right\vert 
+\beta \left(s\right)\left\vert e_0\right\rangle\left\langle e_2\right\vert \right. \nonumber \\
&&\hspace{-0.7cm}\left.+\,\gamma \left( s\right) \left\vert e_3\right\rangle \left\langle
e_1\right\vert +\delta \left( s\right) \left\vert
e_3\right\rangle \left\langle e_2\right\vert + {\textrm{H.C.}} \right] ,
\end{eqnarray}
with H.C. standing for Hermitian conjugate, $\alpha \left( 0\right) =\beta \left( 0\right) =\gamma \left( 0\right)
=\delta \left( 0\right) =1/2$,
and 
\begin{eqnarray}
\hspace{-0.7cm}\alpha \left( 1\right) =\frac{\left( -1\right) ^{f\left( 0\right) +f\left(
1\right) }}{2}, \label{alphaI} \,\,\,
\beta \left( 1\right) =\frac{\left( -1\right) ^{f\left( 0\right) +f\left(
2\right) }}{2}, \label{betaI}\\
\hspace{-0.8cm}\gamma \left( 1\right) =\frac{\left( -1\right) ^{f\left( 1\right) +f\left(
3\right) }}{2}, \label{gammaI} \,\,\,
\delta \left( 1\right) =\frac{\left( -1\right) ^{f\left( 2\right) +f\left(
3\right) }}{2}\label{deltaI}.
\end{eqnarray}
Remarkably, the evolution from $\left\vert \phi\left( 0\right)\right\rangle $ 
to $\left\vert \phi\left( 1\right) \right\rangle$ can be implemented through a 
local Hamiltonian $H\left(s\right)$ on each qubit. Indeed, we propose $H\left(s\right)$ as
\begin{eqnarray}
\hspace{-0.6cm}H\left( s\right) &=&\left( h_{00}^{1}\left\vert 0\right\rangle
_{1}\left\langle 0\right\vert +h_{11}^{1}\left\vert 1\right\rangle
_{1}\left\langle 1\right\vert \right) \otimes \mathbf{1}_{2} \nonumber \\
&&+\mathbf{1}_{1}\otimes \left( h_{00}^{2}\left\vert 0\right\rangle _{2}
\left\langle0\right\vert +h_{11}^{2}\left\vert 1\right\rangle _{2}\left\langle
1\right\vert \right) ,
\end{eqnarray}
where $h_{00}^{i}$ and $h_{11}^{i}$ ($i=1,2$) are real coefficients to be 
determined by the dynamic invariant equation of motion. From 
Eq.~(\ref{dicloseds}), we obtain
\begin{eqnarray}
\hspace{-0.7cm}\alpha \left( s\right) &=&\frac{1}{2}\exp \left\{ iT\int_{0}^{1}\left[
h_{11}^{2}\left( s^{\prime }\right) -h_{00}^{2}\left( s^{\prime }\right) %
\right] ds^{\prime }\right\} , \nonumber \\
\hspace{-0.7cm}\beta \left( s\right) &=&\frac{1}{2}\exp \left\{ iT\int_{0}^{1}\left[
h_{11}^{1}\left( s^{\prime }\right) -h_{00}^{1}\left( s^{\prime }\right) %
\right] ds^{\prime }\right\} , \nonumber \\
\hspace{-0.7cm}\gamma \left( s\right) &=&\beta ^{\ast }\left( s\right), \,\, 
\delta \left( s\right) =\alpha ^{\ast }\left( s\right).
\end{eqnarray}
Since $\left(-1\right) ^{f\left( 0\right) +f\left( 1\right) +f\left( 2\right)
+f\left( 3\right) }=1$, 
Eqs.~(\ref{alphaI})-(\ref{deltaI}) yield $\alpha \left( 1\right)=\delta \left( 1\right)$ 
and $\beta \left( 1\right)=\gamma \left( 1\right)$.
Therefore, a possible simple choice for the Hamiltonian is obtained by defining 
$h_{00}^{1} =-\pi f\left( 2\right)/T$, $h_{11}^{1} =\pi f\left( 0\right)/T$, 
$h_{00}^{2} =-\pi f\left( 1\right)/T$, $h_{11}^{2} =\pi f\left( 0\right)/T$.
Hence, the DJ problem for $n=2$ can be solved by QCDI through the local constant Hamiltonian
\begin{eqnarray}
\hspace{-0.6cm}H &=& \frac{\pi }{2T}\left[ -\left( f\left( 0\right) +f\left( 2\right) \right) \sigma_{z}^{1} \right.\nonumber \\
&&\hspace{-0.9cm}\left.+\left( f\left( 0\right) +f\left( 1\right) \right)\sigma _{z}^{2}
-\left( f\left( 2\right) -f\left( 1\right)\right) \mathbf{1}_{12}\right].
\end{eqnarray}
Observe that no two-body interactions are needed to run the algorithm. For $n>2$, 
interaction terms in $H\left( s\right)$ are expected to appear. Naturally, the scaling of 
such interactions with $n$ is an important issue in order to implement QCDI in large systems.


\subsection{Example 2: The search problem}


A simple implementation of QCDI for the search problem~\cite{Grover:97} can be
given as follows. We start by proposing an oracle in a general form given by 
$U_{0} =\theta|w\rangle\langle w|+\delta\left(  |w\rangle\langle\phi(0)|+
|\phi(0)\rangle\langle w|\right) +\varepsilon|\phi(0)\rangle\langle\phi(0)|$ 
so that $\theta,\delta,\varepsilon\in\mathbb{%
\mathbb{R}
}$ and $\left\vert w\right\rangle $ is the target state in a Hilbert space of $n$ qubits, whose dimension is 
denoted by $N=2^n$. The initial state
$\left\vert \phi(0)\right\rangle $ can be decomposed as $\left\vert
\phi(0)\right\rangle =\alpha\left\vert w\right\rangle +\beta\left\vert
y\right\rangle $, where $\langle w\left\vert y\right\rangle =0$ and
$\alpha^{2}+\beta^{2}=1$, with $\alpha=\left\langle w\right.  \left\vert
\phi(0)\right\rangle $ and $\beta=\left\langle y\right.  \left\vert
\phi(0)\right\rangle $ assumed as real constants. In order to rewrite
$U_{0}$ in terms of $\left\vert w\right\rangle $ and $\left\vert
y\right\rangle $ we define the matrices 
$\mathbf{1}=\left\vert w\right\rangle \left\langle w\right\vert +\left\vert
y\right\rangle \left\langle y\right\vert$, 
$\sigma_{x}=\left\vert w\right\rangle \left\langle y\right\vert +\left\vert
y\right\rangle \left\langle w\right\vert$, 
$\sigma_{y}=-i\left(  \left\vert
w\right\rangle \left\langle y\right\vert -\left\vert y\right\rangle
\left\langle w\right\vert \right)$, and 
$\sigma_{z}=\left\vert
w\right\rangle \left\langle w\right\vert -\left\vert y\right\rangle
\left\langle y\right\vert$.
Then we can write $U_{0}=r_{0}\mathbf{1}+\overrightarrow{r}\cdot
\overrightarrow{\sigma}$, where $r_{0}=\left(  \theta+\varepsilon
+2\alpha\delta\right)  /2$, $\overrightarrow{\sigma}=\left(  \sigma_{x}%
,\sigma_{y},\sigma_{z}\right)  $ and $\overrightarrow{r}=\left(  r_{x}%
,0,r_{z}\right)  $, with $r_{x}=\beta\left(  \delta+\varepsilon\alpha\right)
$ and $r_{z}=\left[  \theta-\varepsilon+2\alpha\left(  \delta+\varepsilon
\alpha\right)  \right]  /2$. Disregarding the term proportional to the
identity, we introduce an interpolation operator $\widetilde{U}(s)$ such as 
in Eq.~(\ref{Interpol1}), with $\alpha(s)=\pi s/2$ and $U$ given by 
$U=(\overrightarrow{r}\cdot\overrightarrow{\sigma})/\left\vert
\overrightarrow{r}\right\vert$.
Note that $U$ is unitary and Hermitian. Let us determine now the conditions
for which $\widetilde{U}(s)$ yields an interpolation between $|\phi(0)\rangle$ 
and the solution state $|w\rangle$. Indeed $\widetilde
{U}(0)=\mathbf{1}$ and therefore $\widetilde{U}(0)|\phi(0)\rangle
=|\phi(0)\rangle$. For the final time we have
\begin{equation}
\widetilde{U}(1)|\phi(0)\rangle=\left[
\frac{\left(  r_{x}\beta+r_{z}\alpha\right)}{i\left\vert \overrightarrow{r}\right\vert} |w\rangle
+\frac{\left(  r_{x}\alpha-r_{z}\beta\right)}{i\left\vert \overrightarrow{r}\right\vert} |y\rangle\right] . 
\nonumber
\end{equation}
Since we want $\widetilde{U}(1)|\phi(0)\rangle=\exp\left(  {i\phi
}\right)  |w\rangle$ (where $\phi$ is an arbitrary unimportant angle), we
impose $r_{x}\alpha=r_{z}\beta$. From this condition, we obtain $\theta
=\varepsilon$. In terms of the vector $\overrightarrow{r}$, this result
implies that $r_{x}=\beta\left(  \delta+\epsilon\alpha\right)$ and 
$r_{z}=\alpha\left(\delta+\epsilon\alpha\right)$. Then 
$\left\vert \overrightarrow
{r}\right\vert =\left\vert \delta+\epsilon\alpha\right\vert$.
Bearing these results in mind, we are then able to build an oracle which
allows for the determination of the element under search at the final time
$s=1$. The dynamic invariant $I(s)$ can be defined by encoding $|\phi(0)\rangle$ 
as its initial ground state and $|w\rangle$ as its final
ground state. Here, this can be conveniently obtained 
by defining $I(0)$ as the projector
$I(0)=\mathbf{1}-\left\vert \phi(0)\right\rangle \left\langle\phi(0)\right\vert$. 
The interpolation given by Eq.~(\ref{Is}) then yields
\begin{align}
\hspace{-0.2cm}I(s)  &  =\mathbf{1} +\frac{i}{2}\sin\left(  \pi s\right)  \left(  |w\rangle\langle\phi(0)|-
|\phi(0)\rangle\langle w|\right) \nonumber\\
&\hspace{-0.6cm} -\cos^{2}\left(  \frac{\pi}{2}s\right)  |\phi(0)
\rangle\langle\phi(0)|-\sin^{2}\left(  \frac{\pi}{2}s\right)
|w\rangle\langle w|  .
\end{align}
Then, the Hamiltonian operator reads
\begin{equation}
\hspace{-0.1cm}H=\frac{\pi}{2T}U=\frac{\pi}{2T}\frac{\overrightarrow{r}\cdot\overrightarrow
{\sigma}}{\left\vert \overrightarrow{r}\right\vert } = 
\pm\frac{\pi}{2T}\left(  \beta\sigma_{x}+\alpha\sigma_{z}\right). \label{Hsearch}%
\end{equation}
In order to have both a dynamic invariant and a Hamiltonian that do not
depend on knowing the value of $|w\rangle$, we fix $\alpha=1/\sqrt{N}$, which
implies $\beta=\sqrt{(N-1)/N}$. Then, Eq.~(\ref{Hsearch}) becomes
\begin{equation}
H=\mp\frac{\pi}{2T\sqrt{N}}\left(  \sqrt{N-1}\sigma_{x}+\sigma_{z}\right)  .
\end{equation}
The equation above resembles the Hamiltonian used in Ref.~\cite{Farhi:98} to
implement the analog analogue of QC. Simulation of such a kind of Hamiltonian
can be implemented by a quantum circuit whose number of oracle calls grows as
$O(\sqrt{N})$.

\vspace{0.5cm}


\section{Conclusion}


\label{SecConc}

In conclusion, we have proposed an approach for implementing QC in continuous time based on 
Lewis-Riesenfeld invariants. Our method opens up the possibility of realizing QC with new 
Hamiltonians and with no adiabatic constraint, 
which may represent a step forward for QC in continuous time (at least for a small-scale regime). 
On the other hand, locality is not ensured {\it a priori} for these Hamiltonians due to 
the unitary interpolation required by the invariant operator. In turn, the absence of the 
adiabaticity condition may be counterbalanced by the hardness of simulating many-body 
interactions in the case of large $n$ QCDI.
Anyway, 
since QC is still in its early stage, the existence of a diversity of QC models can be a
valuable way of providing new experimental routes as well as insights for the
design of quantum algorithms. Robustness of QCDI under decoherence is also an
important further point. In this context, a possible direction can be provided
by the theory of dynamic invariants for open systems as in Ref.~\cite{Sarandy:07}. 
We leave this analysis for future research.

\section*{Acknowledgments}

We gratefully acknowledge financial support from
CNPq, FAPESP, FAPERJ, FAPEMIG, and the Brazilian National Institute for 
Science and Technology of Quantum Information (INCT-IQ).

\end{document}